\documentclass[preprint,10pt,1p,sort&compress]{elsarticle}
\usepackage{amssymb,amsmath,latexsym,mathrsfs,wasysym}
\usepackage{graphicx,subfigure,epsfig}
\usepackage{varioref,xr-hyper,hyperref}
\usepackage{color}
\usepackage{multirow,array,float}
\usepackage[utf8]{inputenc}
\usepackage[T1]{fontenc}

\hypersetup{colorlinks   = true,
            urlcolor     = blue,
            citecolor    = blue,
            linkcolor    = blue,
            menucolor    = blue,
            anchorcolor  = blue,
            filecolor    = blue
            pdfauthor    = author}

\journal{Journal of High Energy Astrophysics}

\begin{document}

\begin{frontmatter}

\title{Implications for the Hubble tension from the ages of the oldest astrophysical objects}

\author[kicc]{Sunny Vagnozzi}\corref{cor1}
\ead{sunny.vagnozzi@ast.cam.ac.uk}\cortext[cor1]{Corresponding author}

\author[cfa,bhi]{Fabio Pacucci}
\ead{fabio.pacucci@cfa.harvard.edu}

\author[cfa,bhi]{Abraham Loeb}
\ead{aloeb@cfa.harvard.edu}

\address[kicc]{Kavli Institute for Cosmology, University of Cambridge, Cambridge CB3 0HA, UK}
\address[cfa]{Center for Astrophysics $\vert$ Harvard \& Smithsonian, Cambridge, MA 02138, USA}
\address[bhi]{Black Hole Initiative, Harvard University, Cambridge, MA 02138, USA}

\begin{abstract}
\noindent We use the ages of old astrophysical objects (OAO) in the redshift range $0 \lesssim z \lesssim 8$ as stringent tests of the late-time cosmic expansion history. Since the age of the Universe at any redshift is inversely proportional to $H_0$, requiring that the Universe be older than the oldest objects it contains at any redshift, provides an upper limit on $H_0$. Using a combination of galaxies imaged from the CANDELS program and various high-$z$ quasars, we construct an age-redshift diagram of $\gtrsim 100$ OAO up to $z \sim 8$. Assuming the $\Lambda$CDM model at late times, we find the 95\%~confidence level upper limit $H_0<73.2\,{\rm km}/{\rm s}/{\rm Mpc}$, in slight disagreement with a host of local $H_0$ measurements. Taken at face value, and assuming that the OAO ages are reliable, this suggests that ultimately a combination of pre- and post-recombination ($z \lesssim 10$) new physics might be required to reconcile cosmic ages with early-time and local $H_0$ measurements. In the context of the Hubble tension, our results motivate the study of either \textit{a)} combined global pre- and post-recombination modifications to $\Lambda$CDM, or \textit{b)} local new physics which only affects the local $H_0$ measurements.
\end{abstract}


\end{frontmatter}

\section{Introduction}
\label{sec:intro}

Historically, the ages of old astrophysical objects (OAO) played an important role in establishing the $\Lambda$CDM cosmological model. Since the 1950s, reports of OAO being ostensibly older than the Universe assuming the prevailing Einstein-de Sitter model~\cite{Dunlop:1996mp,Jimenez:1996at,VandenBerg:1996tm} led to a long-lasting ``age crisis''~\cite{Jaffe:1995qu,1995Natur.376..399B,Krauss:1995yb,Ostriker:1995su,Alcaniz:1999kr}. The discovery of cosmic acceleration through Type Ia Supernovae (SNeIa)~\cite{Riess:1998cb,Perlmutter:1998np} eventually solved this crisis by indicating the need for a dark energy component.

Cosmology now faces another crisis: the Hubble tension, a $\sim 5\sigma$ mismatch between several early-time and local measurements of the Hubble constant $H_0$~\cite{Verde:2019ivm,DiValentino:2021izs,Perivolaropoulos:2021jda,Shah:2021onj,Abdalla:2022yfr}. While systematics cannot yet be excluded as explanation (see e.g.\ Refs.~\cite{Efstathiou:2020wxn,Mortsell:2021nzg,Mortsell:2021tcx,Freedman:2021ahq,Wojtak:2022bct}), if new physics is responsible for the Hubble tension (see e.g.\ Refs.~\cite{Mortsell:2018mfj,Nunes:2018xbm,Poulin:2018zxs,Banihashemi:2018oxo,Guo:2018ans,Poulin:2018cxd,Kreisch:2019yzn,Vattis:2019efj,Lin:2019qug,DiValentino:2019exe,Vagnozzi:2019ezj,DiValentino:2019ffd,Sola:2019jek,Escudero:2019gvw,Sakstein:2019fmf,Akarsu:2019hmw,Ye:2020btb,Yang:2020zuk,Lucca:2020zjb,Barker:2020gcp,Hill:2020osr,Ballesteros:2020sik,Ballardini:2020iws,Gogoi:2020qif,DAmico:2020ods,Gonzalez:2020fdy,Ye:2021nej,Jiang:2021bab,Dialektopoulos:2021wde,Nojiri:2022ski}), the consensus is that it should operate prior to recombination and lower the sound horizon by $\sim 7\%$, to comply with constraints from Baryon Acoustic Oscillations (BAO) and high-$z$ SNe Ia~\cite{Bernal:2016gxb,Addison:2017fdm,Lemos:2018smw,Aylor:2018drw,Knox:2019rjx,Arendse:2019hev,Efstathiou:2021ocp,Schoneberg:2021qvd,Cai:2021weh,Keeley:2022ojz}. However, no compelling model addressing the Hubble tension while keeping a good fit to all available data has been constructed, with recent works casting doubts on the possibility of resolving the Hubble tension through early-time new physics alone~\cite{Krishnan:2020obg,Philcox:2020xbv,Jedamzik:2020zmd,Haridasu:2020pms,Lin:2021sfs,Dainotti:2021pqg,Krishnan:2021dyb,Vagnozzi:2021gjh,Krishnan:2021jmh,Dainotti:2022bzg}.

With a few exceptions~\cite{Jimenez:2003iv,Capozziello:2004jy,Samushia:2009px,Dantas:2010zh,Verde:2013fva,Bengaly:2013afa,Wei:2015cva,Rana:2016gha,Nunes:2020yij,Borghi:2021rft}, OAO received less attention with the end of the ``age crisis''. However, the possibility of shedding light on the Hubble tension via model-independent determinations of the age of the Universe from $z=0$ OAO such as globular clusters and very-low-metallicity stars~\cite{Trenti:2015zja,Jimenez:2019onw,Valcin:2020vav,Bernal:2021yli,Boylan-Kolchin:2021fvy,Krishnan:2021dyb}, has recently been appreciated. Here, we pursue this line of investigation, considering OAO at higher redshifts, up to $z \sim 8$. These allow us to derive an upper limit on $H_0$, while providing a late-time consistency test for $\Lambda$CDM, shedding light on potential ingredients which may help in solving the Hubble tension.

The rest of this paper is then organized as follows. In Sec.~\ref{sec:oao} we explain how OAO can help in arbitrating the Hubble tension. In Sec.~\ref{sec:catalog} we discuss how our OAO catalog is constructed. In Sec.~\ref{sec:analysis} we review our analysis methodology, leading to our results which are presented in Sec.~\ref{sec:results}. A critical discussion of our results is presented in Sec.~\ref{sec:discussion}, and finally we draw concluding remarks in Sec.~\ref{sec:conclusions}.

\section{Ages of old astrophysical objects and the Hubble tension}
\label{sec:oao}

The age-redshift relationship determining the age of the Universe as a function of redshift, $t_U(z)$, is given by:
\begin{eqnarray}
t_U(z) = \int_z^\infty \frac{dz'}{(1+z')H(z')}\,,
\label{eq:tz}
\end{eqnarray}
with $H(z)$ being the Hubble parameter. Eq.~(\ref{eq:tz}) shows that the age of the Universe \textit{at any redshift} is inversely proportional to $H_0 \equiv H(z=0)$, so requiring that the Universe be at least as old as high-$z$ OAO at the appropriate redshifts will lead to \textit{upper} limits on $H_0$ -- if $H_0$ is too high, one ends up in the paradoxical situation of the Universe being younger (at a given redshift) than the oldest objects it contains, as with the pre-1998 ``age crisis''.

In Eq.~(\ref{eq:tz}), $t_U(z)$ picks up most of its contributions at late times ($z \lesssim 10$), and is insensitive to pre-recombination physics. Consistency between high-$z$ upper limits on $H_0$ and local $H_0$ measurements thus provides a test of late-time and/or local new physics, potentially indicating the need for the latter to operate jointly with early-time new physics to fully solve the Hubble tension~\cite{DiValentino:2020srs,Bernal:2021yli,Boylan-Kolchin:2021fvy}.

Different models can predict the same value for the present age of the Universe $t_{U0} \equiv t_U(z=0)$, but a significantly different age-redshift relationship for $z>0$. Therefore, limits on the age of the Universe \textit{at any redshift} can in principle provide more stringent constraints than limits on the \textit{present} age of the Universe $t_{U0}$, or at least help break degeneracies inherent to the latter (see also Refs.~\cite{Wei:2022plg,Moresco:2022phi} for further discussions).

At any redshift $z_i$, the age $t_i$ of an OAO at $z_i$ only sets a \textit{lower} limit to $t_U(z_i) \geq t_i$, since $t_i$ is the difference between the age of the Universe at $z_i$ and when the object was formed at a redshift $z_f$. The reason why the relation between $t_U(z_i)$ and $t_i$ is an inequality is that no object formed at the Big Bang ($z_f \to \infty$). The difference between $t_U(z_i)$ and $t_i$, which we denote by $\tau_{\rm in}$, is sometimes referred to as ``incubation time'', and accounts for the time elapsed between the Big Bang and $z_f$. We expect this to be of order ${\cal O}(0.1)\,{\rm Gyr}$, which is approximately the age of the Universe at $z \sim 20$, when halos of virial temperatures above $10000\,{\rm K}$ could form, allowing the gas to cool by atomic hydrogen transitions and fragment into long-lived stars, thereby allowing for the first generation of low-mass stars to form efficiently (for further discussions on these points, see for example Refs.~\cite{Jimenez:2019onw,Vagnozzi:2020dfn}).

\section{Construction of OAO catalog and age-redshift diagram}
\label{sec:catalog}

We consider galaxies and quasars (QSOs) identified up to $z \approx 8$, with most of our galaxy data coming from the Cosmic Assembly Near-infrared Deep Extragalactic Legacy Survey (CANDELS) observing program~\cite{Grogin:2011ua}. We consider the following observations (their designations in the legend are indicated in parenthesis):
\begin{itemize}
\item CANDELS GOODS-N~\cite{2019ApJS..243...22B}, GOODS-S~\cite{2015ApJ...801...97S}, EGS~\cite{2017ApJS..228....7N}, UDS~\cite{2015ApJ...801...97S}, and COSMOS~\cite{2017ApJS..228....7N} fields (with each specific catalog designated by the respective field name);
\item 32 massive, early-time, passively evolving galaxies in the range $0.12<z<1.85$, with absolute ages determined by Ref.~\cite{Simon:2004tf} (\textit{S05});
\item 7446 QSOs from SDSS DR7, in the range $3 \lesssim z \lesssim 5$~\cite{Shen_2011:DR7} (\textit{SDSS QSOs});
\item 50 QSOs discovered by the GNIRS spectrograph in the range $5.5 \lesssim z \lesssim 6.5$~\cite{Shen:2018ojb} (\textit{Gemini QSOs});
\item 15 QSOs discovered by Pan-STARRS1 in the range $6.5 \lesssim z \lesssim 7.0$~\cite{2017ApJ...849...91M} (\textit{Pan-STARRS1 QSOs});
\item 8 of the most distant QSOs ever detected, in the range $7.0 \lesssim z \lesssim 7.5$~\cite{Banados:2017unc,2019ApJ...872L...2M,Mortlock:2011va,2018ApJ...869L...9W,2019AJ....157..236Y,2019ApJ...883..183M,2020ApJ...897L..14Y}, including J0313-1806, the most distant known QSO~\cite{Wang_2021:QSO}, at redshift $z=7.642$ (\textit{high-z QSOs}).
\end{itemize}
We apply severe quality cuts to the CANDELS galaxies, selecting only those with measured spectroscopic redshifts of secure quality.~\footnote{These cuts amount to only selecting those galaxies with flags \texttt{Spec\_z!=-99.0} and \texttt{Spec\_z\_dq==1} where applicable.} In the following, we only select QSOs with small relative errors on their mass, given the importance of this parameter in robustly inferring the QSOs ages.

\begin{figure}[!t]
\includegraphics[width=0.9\linewidth]{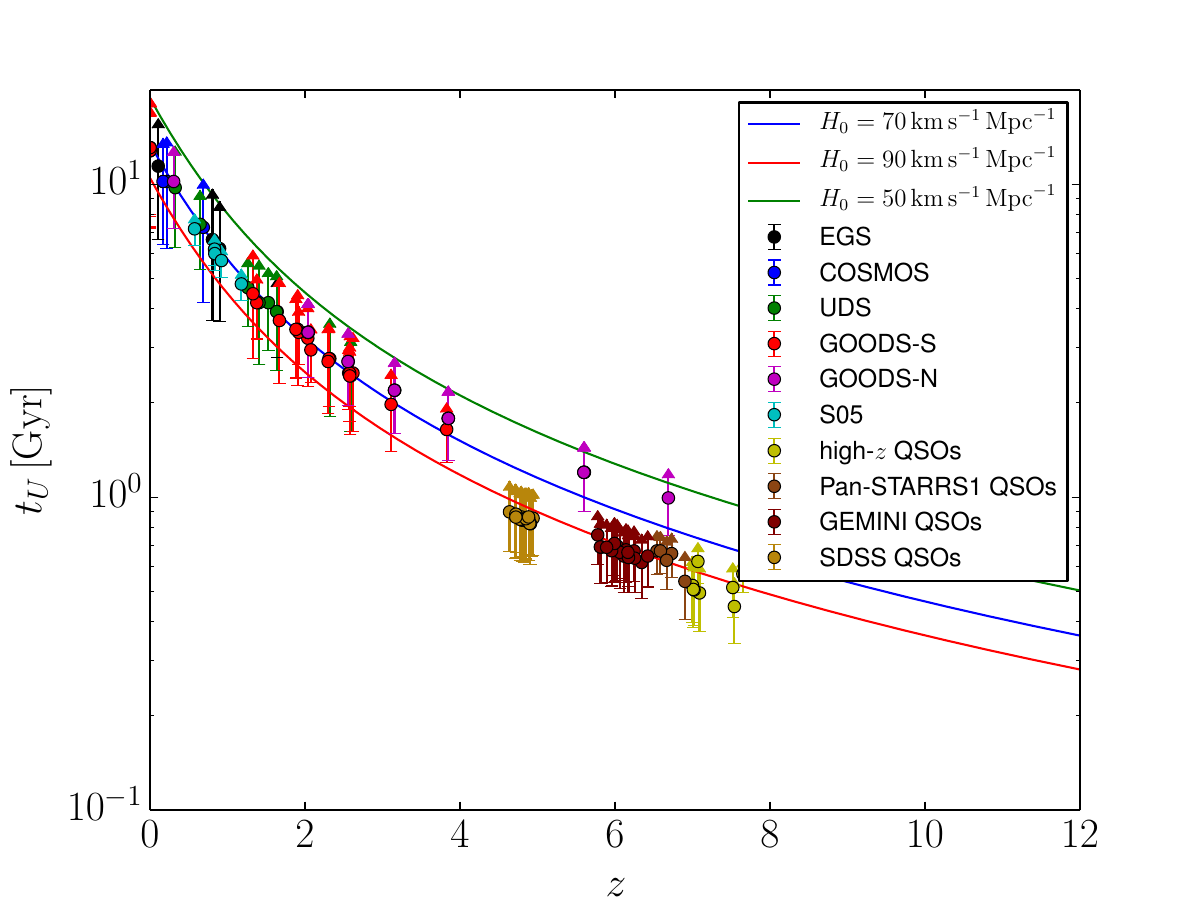}
\caption{Age-redshift diagram for the OAO considered. The curves show the age-redshift relation of $\Lambda$CDM cosmologies with $\Omega_m=0.3$ and $H_0$ determined by the color coding.}
\label{fig:ooa}
\end{figure}

The ages of the CANDELS galaxies have been estimated via a wide variety of methods through the conjoined efforts of 10 teams~\cite{2015ApJ...801...97S,2019ApJS..243...22B}. The general methodology relies on fitting the photometric spectral energy distributions with spectral population synthesis models, with further assumptions on dust attenuation, star formation history, and galactic luminosities and colors~\cite{2019ApJS..243...22B}. While age estimates across different methods agree well with each other, for consistency we adopt the $6a_{\tau}^{\rm NEB}$ method~\cite{2006A&A...459..745F}, one of the few to have included nebular emission and continuum in the model spectra, leading to more robust results. We conservatively estimate galaxy ages uncertainties via the standard deviation of the ages across different methods, finding that the relative uncertainties typically fall between $25\%$ and $40\%$. For \textit{S05}, we follow Ref.~\cite{Simon:2004tf} and use $12\%$ relative uncertainties.

To estimate QSOs ages and uncertainties, we employ the integration methods described in Sec.~3 of Ref.~\cite{Pacucci:2017mcu}. The target function to compute the age $t_{\rm obs}$ of the QSO at the observation redshift $z_{\rm obs}$ is:
\begin{eqnarray}
t_{\rm obs} = \frac{t_S}{{\cal D} f_{\rm Edd}} \frac{\epsilon}{1-\epsilon} \ln\frac{M_{\bullet}}{M_{\rm seed}} \,
\end{eqnarray}
with $t_S \sim 0.45\,{\rm Gyr}$ being the Salpeter time~\cite{Salpeter_1964}, ${\cal D}$ the duty cycle, $f_{\rm Edd}$ the Eddington ratio, $\epsilon$ the radiative efficiency, $M_{\bullet}$ and $M_{\rm seed}$ the observed mass and seed mass. We assume a bimodal prior for $M_{\rm seed}$ following Ref.~\cite{Pacucci_2019dec}, and a flat prior for ${\cal D} \in [0;1]$. For $f_{\rm Edd}$, we assume a Gaussian prior based on observational determinations where available (conservatively setting the standard deviation to $0.5$ if not available), or else a flat prior $f_{\rm Edd} \in [0;2]$. We assume $\epsilon \sim 6\%$, in agreement with simulations~\cite{Jiang_2019}, and a seeding redshift $z_f \sim 20$~\cite{BL_2001} (with our results being only weakly sensitive to this choice). For age determination purposes, our choices regarding $\epsilon$ and $z_f$ are conservative. For each source, we compute $10^4$ Monte Carlo realizations of the growth model and infer the QSOs ages and uncertainties from the resulting statistical distributions of these parameters.
 
We divide each catalog into redshift bins, and select only those objects which are among the oldest ones within each bin. The previous quality cut and this ``age cut'' leave us with 114 OAO, whose age-redshift diagram is given in Fig.~\ref{fig:ooa}. To the best of our knowledge, it is the first time such an extensive high-$z$ OAO catalog and associated age-redshift diagram have been compiled.

\section{Analysis methodology}
\label{sec:analysis}

We consider a model described by 3 parameters: $\boldsymbol{\theta} \equiv \{\Omega_m$, $H_0$, $\tau_{\rm in}\}$ (with $\Omega_m$ the matter density). We perform a Bayesian analysis to constrain these parameters against the OAO ages, making use of Markov Chain Monte Carlo methods, with chains generated via the \texttt{MontePython3.3}~\cite{Audren:2012wb,Brinckmann:2018cvx} sampler.

We assume that the late-time expansion history is described by $\Lambda$CDM, in such a way that the Hubble rate in Eq.~(\ref{eq:tz}) can be well approximated by $H(z) \approx H_0\sqrt{\Omega_m(1+z)^3+(1-\Omega_m)}$. This then allows for a valuable consistency test: if we trust the OAO ages, a disagreement between our upper limit on $H_0$ and the locally measured value(s) would be an indication for missing beyond-$\Lambda$CDM ingredients \textit{at least} in the late-time expansion history. This conclusion would be independent of any assumption on the pre-recombination expansion, given its negligible contribution to $t_U(z)$.

Consider a data vector $\boldsymbol{d} \equiv \{z_i,t_i,\sigma_{t_i} \}$, with the OAO ages at redshifts $z_i$ being $t_i \pm \sigma_{t_i}$. The probability of observing $\boldsymbol{d}$ given a choice of parameters $\boldsymbol{\hat{\theta}}$ is modelled through the following half-Gaussian (log-)likelihood:
\begin{eqnarray}
\ln{\cal L}(\boldsymbol{\hat{\theta}} \vert \boldsymbol{d}) = -\frac{1}{2} \sum_i
\begin{cases}
\Delta_i^2(\boldsymbol{\hat{\theta}})/\sigma_{t_i}^2 & \text{if}\ \Delta_i(\boldsymbol{\hat{\theta}})<0 \\
0 & \text{if}\ \Delta_i(\boldsymbol{\hat{\theta}}) \geq 0
\end{cases}
\,,
\label{eq:loglikelihood}
\end{eqnarray}
where $\Delta_i \equiv t_U(\boldsymbol{\hat{\theta}},z_i)-t_i-\tau_{\rm in}$. Considering the $i$th OAO, Eq.~(\ref{eq:loglikelihood}) is expressing the fact that: \textit{a)} parameters for which the Universe is older than the OAO plus the incubation time (\textit{i.e.} $\Delta_i(\boldsymbol{\hat{\theta}}) \geq 0$) are equally likely, and cannot be distinguished based on the OAO age alone; \textit{b)} parameters for which the Universe is younger than the OAO plus the incubation time (\textit{i.e.} $\Delta_i(\boldsymbol{\hat{\theta}}) < 0$) are (exponentially) unlikely, as the Universe clearly cannot be younger than its oldest inhabitants.

In our baseline analysis, we set flat priors on $H_0 \in [40;100]\,{\rm km}/{\rm s}/{\rm Mpc}$ and $\Omega_m \in [0.2;0.4]$. We choose wide prior ranges for $H_0$ and $\Omega_m$ in order to be as conservative as possible, but have verified that our results are only weakly sensitive to the choice of prior boundaries (see also Ref.~\cite{Wei:2022plg}).

We model $\tau_{\rm in}$ following Ref.~\cite{Jimenez:2019onw}, which derives a probability distribution for $\tau_{\rm in}$ from a set of reasonable empirical assumptions on the formation redshift of the oldest galaxies $z_f$. After marginalizing over $H_0$, $\Omega_m$, and $z_f$, a prior peaked around $\tau_{\rm in} \approx 0.1-0.15\,{\rm Gyr}$ is found.~\footnote{This reflects the time when halos of virial temperatures above $10^4\,{\rm K}$ formed, allowing gas to be cooled by atomic hydrogen transitions and fragment into long-lived stars, enabling the first generation of low-mass stars to form efficiently.} We refer to this prior as J19 (from the initial of the first author of Ref.~\cite{Jimenez:2019onw}), and use its fitting function provided in Appendix~G of Ref.~\cite{Valcin:2020vav}. The J19 prior is very conservative in the sense that it identifies our OAO as being descendants of the oldest generation of galaxies. Assuming they instead descended from galaxies at lower redshifts would push the $\tau_{\rm in}$ distribution towards higher values. From Eqs.~(\ref{eq:tz},\ref{eq:loglikelihood}), this would clearly result in more stringent, yet less conservative, limits on $H_0$. For the QSOs, we fix $\tau_{\rm in}=t_U(z_f=20)$, given our assumption of a seeding redshift $z_f \sim 20$.

\section{Results}
\label{sec:results}

\begin{figure}[!bth]
\includegraphics[width=0.9\linewidth]{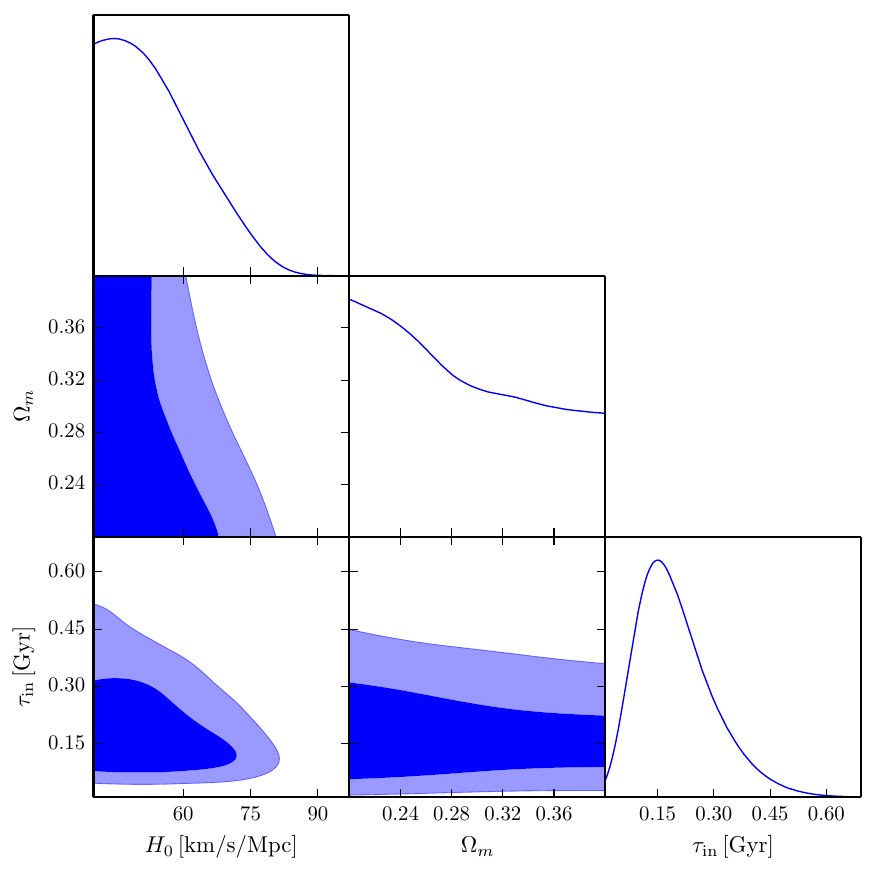}
\caption{2D joint and 1D marginalized posterior distributions for $H_0$, $\Omega_m$, and the incubation time $\tau_{\rm in}$, from our baseline analysis of the ages of the objects shown in Fig.~\ref{fig:ooa}. For $H_0$ we find the 95\%~C.L. upper limit $H_0<73.2\,{\rm km}/{\rm s}/{\rm Mpc}$.}
\label{fig:base}
\end{figure}

The corner plot in Fig.~\ref{fig:base} shows the joint $H_0$-$\Omega_m$-$\tau_{\rm in}$ posterior distributions obtained from our baseline analysis, with flat priors on $H_0$ and $\Omega_m$, and the J19 prior on $\tau_{\rm in}$. The $\tau_{\rm in}$ posterior unsurprisingly follows the J19 prior, whereas $\Omega_m$ is unconstrained, and we can set an upper limit on $H_0$ as expected. Our 95\% confidence level (C.L.) upper limit is $h_0<0.732$ (hereafter, all quoted upper limits will be at 95\%~C.L.), with $h_0 \equiv H_0/(100\,{\rm km}/{\rm s}/{\rm Mpc})$ being the reduced Hubble constant. We regard this upper limit as being the most balanced one in terms of equilibrium between conservative and aggressive assumptions.

The upper limit $h_0<0.732$ is only marginally consistent with various local measurements of $H_0$. For instance, the distance ladder constructed out of Cepheid-calibrated SNe Ia finds $h_0=0.7403 \pm 0.0142$~\cite{Riess:2019cxk} (R19 hereafter). Using the full non-Gaussian information from our $H_0$ posterior, we quantify the concordance/discordance between our inferred $H_0$ and R19 as being at the $2.3\sigma$ level. While falling in a grey zone which does not fully qualify as a tension, it is still a discrepancy worthy of attention.

We investigate how our results are affected by other choices of priors. To explore the impact of the J19 prior, we fix $\tau_{\rm in}$ to $0.1\,{\rm Gyr}$ and $0\,{\rm Gyr}$. In the former case we find $h_0<0.762$, which relaxes to $h_0<0.791$ in the latter, in both cases relaxing the discrepancy with the R19 measurement. However, both values are overly conservative, with the second being completely unrealistic as $\tau_{\rm in}=0\,{\rm Gyr}$ amounts to the assumption that the OAO formed at the Big Bang.

We then set a Gaussian prior $\Omega_m=0.315 \pm 0.007$, informed by the \textit{Planck} 2018 constraints within the $\Lambda$CDM model~\cite{Aghanim:2018eyx}, finding $h_0<0.674$, in $2.5\sigma$ tension with R19. We caution against over-interpreting this result, as it relies on a very tight external prior which assumes a specific cosmological model for both the late and, importantly, early Universe. A less aggressive approach could entail adopting a more generous (wider) prior on $\Omega_m$. For example, setting a Gaussian prior $\Omega_m=0.30 \pm 0.05$, we find an upper limit of $h_0<0.714$. This prior is not informed by any specific probe, but reflects the fact that a wide range of independent and robust late-time probes point towards $\Omega_m \approx 0.3$ (see for instance Refs.~\cite{Moresco:2016mzx,Vagnozzi:2017ovm,Abbott:2017wau,Zubeldia:2019brr,Asgari:2020wuj}).

Finally, we set a Gaussian prior on $h_0=0.7403 \pm 0.0142$, informed by the R19 measurement, and find $\Omega_m<0.29$. An upper limit on $\Omega_m$ is to be expected since, at fixed $H_0$, increasing $\Omega_m$ decreases $t(z)$ for all $z$: thus, increasing $\Omega_m$ sufficiently would make the Universe too young (at any $z$) to accommodate the OAO ages, explaining why we find an upper limit on $\Omega_m$.

\section{Discussion}
\label{sec:discussion}

Our baseline upper limit $h_0<0.732$ appears to indicate a discrepancy with a few local measurements, especially the R19 Cepheid-calibrated SNe Ia one~\cite{Riess:2019cxk}. Our limit hinges upon two assumptions: \textit{i)} the reliability of the adopted OAO ages, and \textit{ii)} the validity of the $\Lambda$CDM model at low redshifts. Concerning \textit{ii)}, we make no assumptions on the pre-recombination expansion history, nor would our results be sensitive to these, as they essentialy only depend on the assumed expansion history for $z \lesssim 10$.

Taken at face value, this appears to indicate that while $\Lambda$CDM remains an excellent fit to late-time observations such as BAO~\cite{Beutler:2011hx,Ross:2014qpa,Alam:2020sor} and high-$z$ SNe Ia~\cite{Scolnic:2017caz}, a small but non-negligible amount of global late-time ($z \lesssim 10$) new physics might be required to improve the agreement between OAO ages and local $H_0$ measurements. At $z>0$, this needs to go in the direction of lowering the expansion rate relative to $\Lambda$CDM to accommodate a higher $H_0$: an example is a phantom dark energy component (see for instance Refs.~\cite{Renk:2017rzu,Vagnozzi:2018jhn,Yang:2018euj,Li:2019yem,Visinelli:2019qqu,Pan:2019hac,DiValentino:2019jae,Zumalacarregui:2020cjh,Alestas:2020mvb,Banihashemi:2020wtb,Bag:2021cqm,Alestas:2021luu,Chudaykin:2022rnl,Sharma:2022ifr,Nunes:2022bhn} for relevant discussions on phantom dark energy components in the context of the Hubble tension).

This conclusion does not contradict the earlier ones of Refs.~\cite{Bernal:2016gxb,Addison:2017fdm,Lemos:2018smw,Aylor:2018drw,Knox:2019rjx,Arendse:2019hev,Efstathiou:2021ocp,Schoneberg:2021qvd,Cai:2021weh,Keeley:2022ojz}, indicating that global late-time new physics \textit{alone} cannot resolve the Hubble tension: if global new physics is responsible for the Hubble tension, the majority of it will unquestionably have to operate at early times. However, there still is room for early- and late-time new physics working in the same direction to potentially reduce the Hubble tension to a statistically acceptable level. These results are broadly in line with Refs.~\cite{Krishnan:2020obg,Philcox:2020xbv,Jedamzik:2020zmd,Haridasu:2020pms,Lin:2021sfs,Dainotti:2021pqg,Krishnan:2021dyb,Vagnozzi:2021gjh,Krishnan:2021jmh,Dainotti:2022bzg} which, while approaching the problem from a wide range of different perspectives, also highlighted the difficulty in resolving the Hubble tension with early-time new physics \textit{alone}.

It was recently pointed out that, if a high $t_{U0}$ were measured reliably and to high precision, a combination of global pre- and post-recombination new physics, or local new physics, would be needed to reconcile all measurements and solve the Hubble tension~\cite{DiValentino:2020srs,Bernal:2021yli}. Our results, focusing on $t_U(z)$ rather than $t_{U0}$ alone, may be the first hint of this situation. Related $t_{U0}$-based hints were also recently reported~\cite{Krishnan:2021dyb}.

While BAO and high-$z$ SNe Ia measurements set tight constraints on late-time deviations from $\Lambda$CDM, there is within the error bars still some wiggle room to accommodate late-time new physics to improve the agreement between OAO ages and local $H_0$ measurements. Combining \textit{Planck} 2018 and BAO measurements constrains the dark energy equation of state to $w=-1.04^{+0.06}_{-0.05}$~\cite{Aghanim:2018eyx}, while \textit{Planck}+\textit{Pantheon} SNe Ia gives $w=-1.03 \pm 0.04$~\cite{Scolnic:2017caz}, with looser limits obtained relaxing assumptions on curvature~\cite{DiValentino:2020hov}. Within these limits there is most certainly still wiggle room in the $w<-1$ direction for new late-time physics operating together with new early-time physics to improve the agreement between OAO and local measurements of $H_0$, and reduce the Hubble tension.

Our results hinge upon the trustworthiness of the OAO ages. It is of course no easy task to estimate galaxies and QSOs ages at high-$z$, and these are subject to several systematic uncertainties~\cite{Valcin:2021jcg}. Despite our conservative approach, we cannot exclude that the discrepancy between OAO ages and local $H_0$ measurements could be due to the former being overestimated (or the latter underestimated). The potential detection, with upcoming facilities~\cite{Pacucci_2019dec,Inayoshi_2020}, of farther galaxies and QSOs would significantly improve our ability to estimate their ages. Detecting a QSO at $z\gtrsim 10$ would dramatically shrink the available growth parameters phase space, leading to a more accurate age estimate (see Ref.~\cite{Pacucci:2021ubg}).

\section{Conclusions}
\label{sec:conclusions}

We have shown the potential of the ages of high-$z$ old astrophysical objects (OAO) to constrain the late-time expansion history and arbitrate the Hubble tension. This method's strength rests upon its not making assumptions on the pre-recombination expansion history while providing an upper limit on $H_0$, since raising $H_0$ makes the Universe younger at any $z$, potentially bringing us into the paradoxical situation where the Universe is younger than these OAO -- a few decades ago, a similar situation was indeed providing the first hints for the existence of dark energy~\cite{Krauss:1995yb,Ostriker:1995su}, and we have argued that the time may now be ripe for OAO to once again play an important role for cosmology in the context of the Hubble tension.

Compiling an age-redshift diagram of OAO up to $z \sim 8$ using galaxies and QSOs we find, assuming the validity of $\Lambda$CDM at late times, $H_0<73.2\,{\rm km}/{\rm s}/{\rm Mpc}$ at 95\%~C.L., discrepant with local $H_0$ measurements. This suggests at face value that a small but non-negligible amount of late-time ($z \lesssim 10$) new physics might be required to alleviate this discrepancy. This, in combination with early-time new physics lowering the sound horizon at recombination, could help bringing down the Hubble tension to an acceptable level. In addition, it is tempting to speculate that this same late-time new physics may also play a role in alleviating the $S_8$ tension~\cite{DiValentino:2018gcu,DiValentino:2020vvd,Nunes:2021ipq}: see for instance Refs.~\cite{Alestas:2021xes,Allali:2021azp,Ye:2021iwa,Anchordoqui:2021gji,Khosravi:2021csn,Clark:2021hlo,Heisenberg:2022lob,Heisenberg:2022gqk,Davari:2022uwd,Reeves:2022aoi} for recent relevant discussions.

Our results indicate two important research directions. Firstly, improving the reliability of  high-$z$ astrophysical objects age determinations is an important priority~\cite{Borghi:2021zsr}, and the recent launch of JWST will allow for significant progress along these lines. Moreover, it is worth exploring the possibility of addressing the Hubble tension via a combination of global late-time and (mostly) early-time new physics, thus far mostly considered separately: our results suggest that a combination of the two might ultimately be necessary, broadly supporting recent findings~\cite{Krishnan:2020obg,Philcox:2020xbv,Jedamzik:2020zmd,Haridasu:2020pms,Lin:2021sfs,Dainotti:2021pqg,Krishnan:2021dyb,Vagnozzi:2021gjh,Krishnan:2021jmh,Dainotti:2022bzg}. Valid alternatives would be to invoke local new physics only affecting local $H_0$ measurements~\cite{Lombriser:2019ahl,Desmond:2019ygn,Ding:2019mmw,Desmond:2020wep,Alestas:2020zol,Cai:2021wgv,Marra:2021fvf}, or a breakdown of the FLRW framework~\cite{Secrest:2020has,Krishnan:2020vaf,Krishnan:2021dyb,Krishnan:2021jmh,Aluri:2022hzs}. Our overall message is that cosmic ages are an extremely valuable tool in the quest towards unraveling the Hubble tension (see also Refs.~\cite{Wei:2022plg,Moresco:2022phi}).

\section*{Acknowledgements}
\noindent S.V. acknowledges very useful discussions with George Efstathiou, Raul Jim\'{e}nez, and Michele Moresco. S.V. is supported by the Isaac Newton Trust and the Kavli Foundation through a Newton-Kavli Fellowship, and by a grant from the Foundation Blanceflor Boncompagni Ludovisi, n\'{e}e Bildt. S.V. acknowledges a College Research Associateship at Homerton College, University of Cambridge. F.P. acknowledges support from a Clay Fellowship administered by the Smithsonian Astrophysical Observatory. A.L. and F.P. are partially supported by the Black Hole Initiative at Harvard University, which is funded by grants from the John Templeton Foundation (JTF) and the Gordon and Betty Moore Foundation (GBMF). This work was performed using resources provided by the Cambridge Service for Data Driven Discovery (CSD3) operated by the University of Cambridge Research Computing Service (\href{https://www.hpc.cam.ac.uk/}{www.hpc.cam.ac.uk}), provided by Dell EMC and Intel using Tier-2 funding from the Engineering and Physical Sciences Research Council (capital grant EP/P020259/1), and DiRAC funding from the Science and Technology Facilities Council (\href{https://www.dirac.ac.uk/}{www.dirac.ac.uk}).

\footnotesize

\bibliography{Cosmology_OAO_JHEAp.bib}
\bibliographystyle{JHEP}

\end{document}